\documentclass[trackchanges]{aastex701}

\begin{document}

\title{JWST/MIRI Observations of Amorphous Silicates in Interstellar Comet 3I/ATLAS}

\correspondingauthor{C. A. Thomas}
\email{cristina.thomas@nau.edu}

\author[0000-0003-3091-5757]{Cristina A. Thomas}
\affiliation{Northern Arizona University, Department of Astronomy and Planetary Science,
P.O. Box 6010,
Flagstaff, AZ, 86011 USA}
\email{cristina.thomas@nau.edu}

\author[0000-0002-6702-7676]{Michael S. P. Kelley}
\affiliation{Department of Astronomy, University of Maryland, College Park, MD 20742-0001, USA}
\email{msk@astro.umd.edu}

\author[0000-0001-6397-9082]{David E. Harker}
\affiliation{Department of Astronomy and Astrophysics, University of California, San Diego, 9500 Gilman Drive, MC 0424, La Jolla, CA 92093-0424, USA}
\email{dharker@ucsd.edu}

\author[0000-0001-8233-2436]{Martin A. Cordiner}
\affiliation{Astrochemistry Laboratory, NASA Goddard Space Flight Center, 8800 Greenbelt Road, Greenbelt, MD 20771, USA.}
\affiliation{Department of Physics, Catholic University of America, Washington, DC 20064, USA.}
\email{martin.cordiner@nasa.gov}

\author[0000-0003-0774-884X]{Davide Farnocchia}
\affiliation{Jet Propulsion Laboratory, California Institute of Technology, 4800 Oak Grove Dr., Pasadena, CA 91109, USA.}
\email{Davide.Farnocchia@jpl.nasa.gov}

\author[0000-0002-2668-7248]{Dennis Bodewits}
\affiliation{Physics Department, Edmund C. Leach Science Center, Auburn University, Auburn, AL 36849, USA.}
\email{dennis@auburn.edu}

\author[0000-0001-6752-5109]{Steven B. Charnley}
\affiliation{Astrochemistry Laboratory, NASA Goddard Space Flight Center, 8800 Greenbelt Road, Greenbelt, MD 20771, USA.}
\email{steven.b.charnley@nasa.gov}

\author[0000-0001-7479-4948]{Maria N. Drozdovskaya}
\affiliation{Department of Chemistry, Biochemistry and Pharmaceutical Sciences (DCBP), Universit¨at Bern, Freiestrasse 3, 3012 Bern, Switzerland}
\email{maria.drozdovskaya.space@gmail.com}

\author[0000-0001-7895-8209]{Marco Micheli}
\affiliation{ESA NEO Coordination Centre, Planetary Defence Office, European Space Agency, Largo Galileo Galilei, 1, 00044 Frascati (RM), Italy.}
\email{marco.bs.it@gmail.com}

\author[0000-0001-7694-4129]{Stefanie N. Milam}
\affiliation{Astrochemistry Laboratory, NASA Goddard Space Flight Center, 8800 Greenbelt Road, Greenbelt, MD 20771, USA.}
\email{stefanie.n.milam@nasa.gov}
\author[0000-0001-7694-4129]{Cyrielle Opitom}
\affiliation{Institute for Astronomy, University of Edinburgh, Royal Observatory, Edinburgh EH9 3HJ, UK.}
\email{copi@roe.ac.uk}

\author[0000-0001-8541-8550]{Silvia Protopapa}
\affiliation{Solar System Science and Exploration Division, Southwest Research Institute, Boulder, CO 80302, USA}
\email[show]{silvia.protopapa@swri.org}

\author[0000-0002-6006-9574]{Nathan X. Roth}
\affiliation{Astrochemistry Laboratory, NASA Goddard Space Flight Center, 8800 Greenbelt Road, Greenbelt, MD 20771, USA.}
\affiliation{Department of Physics, American University, 4400 Massachusetts Avenue, NW, Washington, DC 20016, USA.}
\email{nathaniel.x.roth@nasa.gov}

\author[0000-0003-4365-1455]{Megan E. Schwamb}
\affiliation{Astrophysics Research Centre, School of Mathematics and Physics, Queen's University Belfast, Belfast BT7 1NN, UK.}
\email{m.schwamb@qub.ac.uk}

\begin{abstract}

We present JWST/MIRI Low Resolution Spectroscopy (LRS) of 3I/ATLAS obtained post-perihelion at a heliocentric distance of 2.21 au. Our spectrum contains a prominent 10 $\mu$m silicate emission feature with a broad, smooth shape that is distinct from those commonly observed in Solar System comets. We model the spectrum using thermal emission from amorphous carbon, amorphous olivine, and amorphous pyroxene grains. Our models strongly favor a pyroxene-dominated, amorphous silicate composition and show a degeneracy between silicate grain size and the relative abundance of silicate and carbon dust. Across a set of representative models, the best-fitting crystalline abundance is zero, and we place an upper limit of 5.3\% on the crystalline fraction of the silicates. We identify an emission feature near 6.83 $\mu$m that is consistent with aliphatic C–H-bearing material. The overwhelmingly amorphous silicate composition supports previous work that measured unusual volatile and isotopic composition in suggesting that the material incorporated into the object experienced little thermal processing and preserves information about the conditions in which this interstellar planetesimal formed.

\end{abstract}

\keywords{\uat{Interstellar objects}{52} --- \uat{James Webb Space Telescope}{2291} --- \uat{Infrared spectroscopy}{2285} --- \uat{Comets}{280}}


\section{Introduction} 
Interstellar objects (ISOs) are planetesimals formed in planetary systems far from our own, under different chemical and physical conditions. An ISO's passage through the Solar System presents a rare opportunity to observe extrasolar material at close enough range for detailed spectroscopic analyses \citep[e.g.,][]{fitzsimmons2019detection,fitzsimmons2018spectroscopy,ye20171i,opitom20192i}. Measurements of their activity, component volatiles, and refractory dust compositions can provide direct constraints on the diversity of planetesimal formation and processing in other planetary systems.

The interstellar comet 3I/ATLAS is the third documented interstellar object following the discoveries of 1I/'Oumuamua \citep[e.g.,][]{meech2017brief} and 2I/Borisov \citep[e.g.,][]{jewitt2019initial} in 2017 and 2019, respectively. The first two interstellar objects presented fundamentally different attributes: 1I showed no coma or dust tail, while 2I was actively producing gas and dust at discovery. This contrast suggested that future ISO observations should be designed to characterize both inactive and cometary objects. Our JWST Target-of-Opportunity program was designed with the goal to quickly obtain a wide range of diagnostic observations considering a range of possible activity states.

Several recent studies conclude that 3I/ATLAS formed under different conditions than those of the Solar System comets. 
JWST NIRSpec observations of the gas coma \citep{cordiner2026isotopic} found exceptionally high D/H and $^{12}$C/$^{13}$C ratios. ALMA and VLT observations independently observed the elevated carbon isotope ratio \citep{salazar2026water,opitom2026high}. The VLT data also showed $^{14}$N/$^{15}$N to be significantly higher than typically observed in Solar System comets\citep{opitom2026high}. Other notable differences from Solar System comets include high initial Fe, Ni, and CO$_2$ abundances and relatively low overall outflow velocities consistent with CO/CO$_2$-dominated outgassing \citep{biver2026perihelion}. 
All together, these measurements are direct evidence of the conditions under which the object formed and imply formation temperatures $\leq$30~K in a relatively metal-poor environment.  

We present observations of 3I/ATLAS using the James Webb Space Telescope \citep[JWST, e.g.,][]{rigby2023science, Gardner_2023} Mid-Infrared Instrument (MIRI) Low Resolution Spectroscopy (LRS) mode. This spectrum was taken as part of a target-of-opportunity program that also included NIRSpec IFU spectroscopy \citep{cordiner2025jwst,cordiner2026isotopic} and imaging with MIRI and NIRCam \citep{kelley2026imaging}. In this paper, we investigate the dust composition of 3I's coma with a focus on the relative contributions of amorphous pyroxene, amorphous olivine, and carbon. We examine how the silicate composition differs from that of Solar System comets and what the lack of detectable crystalline silicates reveals about its thermal processing history. These results provide an independent constraint on the history of 3I’s refractory material, complementing previous measurements of its volatile and isotopic composition.

\section{Observations \& Reduction} \label{sec:obs_reduc}

3I/ATLAS was observed with the JWST Mid-Infrared Instrument \citep[MIRI,][]{wright2023mid} low-resolution spectrometer \citep{kendrew2015mid} in fixed slit (mode as part of a series of observations in program ID 5094. This spectral mode uses a 4\farcs7 long and 0\farcs51 wide slit to provide data from 5--14 $\mu$m with spectral resolving power of $\sim$100 at 7.5 $\mu$m. We performed target acquisition and pointing verification using the F560W filter. 

Our observation sequence consisted of two integrations dithered along the slit with two complementary background observations offset $\sim$28\farcm2 away from the coordinates of 3I. Each integration had an effective exposure time of 224.8 seconds with an effective total exposure time of 449.6 seconds. We used the F560W filter for target acquisition and the P750L dispersing element with the FASTR1 readout pattern for spectroscopy. The science sequence began at UT 2025 December 16 16:37:59 when 3I was at a (post-perihelion) heliocentric distance $r=2.214$ au and comet-JWST distance $\Delta=1.798$ au. The background sequence was taken hours later and started at UT 2025 December 17 02:01:56. The JWST data presented here were obtained from the Mikulski Archive for Space Telescopes (MAST) at the Space Telescope Science Institute. The specific observations analyzed can
be accessed via DOI: 10.17909/1mye-qv08. 

\begin{figure*}[ht!]
\plotone{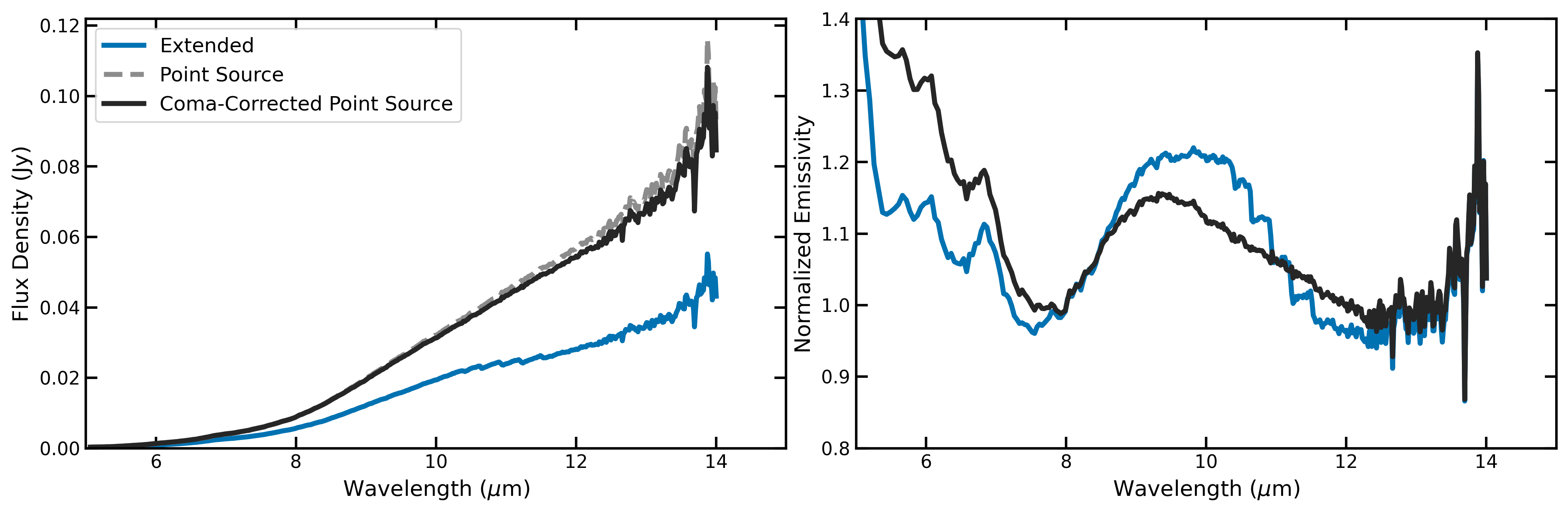}
\caption{Left: Extracted 3I spectral flux using the LRS pipeline point and extended settings compared to our coma-corrected point source method. Right: A comparison of the normalized emissivity spectrum using the extended source pipeline products and our coma-corrected point source method. The extended source extraction shows a fundamentally different shape for the 10 $\mu$m emissivity feature including significant artifacts around 10-12 $\mu$m. As a result, the spectrum from the extended-source extraction cannot reliably characterize the dust mineralogy.}
\label{fig:spectra}
\end{figure*}

The MIRI-LRS data were reduced using the JWST Calibration Pipeline software version v1.20.2 \citep{bushouse2024} with the Calibration Reference Data System context file 1464. 
3I was active and extended at the time of observation, and clear artifacts in the pipeline-reduced data show that the nominal data reduction process for extended objects did not accurately recover the LRS spectrum. The extended object reduction process assumes uniform slit illumination and does not apply a correction for wavelength dependent slit losses. 3I's spatial structure is intermediate between an extended object and a point source. Additionally, the extended object process did not account for significant artifacts in the 10--12~\micron{} region, which results in a sawtooth pattern that alters the shape of the silicate emissivity feature. These artifacts are accounted for in the point source reduction process.  A comparison of the extended and point source processing is shown in Fig.~\ref{fig:spectra}.
To address these calibration issues, our reduction used the point-source pipeline process with additional corrections for the extended coma.

The 3I exposures and the off-source background observations were linked in an association file. The spec2pipeline was run with the source type set to ``point" to apply the pathloss (diffraction loss) and aperture loss corrections. Our resulting 3I spectra were combined using inverse-variance weighting to produce the final spectrum. We subsequently applied a wavelength-dependent throughput correction to the extracted spectrum to account for 3I’s extended spatial profile and the resulting differences in slit and aperture throughput relative to a true point source \citep[e.g.,][]{kelley2006spitzer}. The process used MIRI LRS PSF models generated by the STPSF software \citep{perrin2012stpsf} for wavelengths between 5.0 and 14.0~\micron{} (0.1~\micron{} steps, 4$\times$ spatial oversampling).  A nominal cometary coma was generated with a power-law surface brightness distribution as a function of distance to the (unresolved) nucleus: $I\propto1/\rho$.  The total brightness within an aperture of dimensions 4.6$\times$8~pix (corresponding to the slit width and extraction aperture) was calculated for the simulated coma, and the coma convolved with the PSFs.  The total correction for the pathloss and aperture is computed from the ratio of the two totals.  All spectra (Figure \ref{fig:spectra}, left) were trimmed to wavelengths $\geq$ 5.2 µm for our analysis to exclude the low signal-to-noise ratio data at the shortest wavelengths. The formal uncertainties on the remainder of the extracted spectrum are extremely small, with a median uncertainty of $\sim$0.13\%. This quoted uncertainty only represents the errors of the extracted point-source spectrum. 

Our reduced spectrum is generally consistent with the spectrum from the near-contemporaneous MIRI MRS observation \citep[DDT, program ID 9442,][]{belyakov2026volatile,belyakov2026dust} obtained less than a day earlier. The shape of the silicate emissivity feature is similar, but there is a difference in spectral contrast that may be due to spectral calibration, comet variability, or a combination of both. While the emissivity feature resulting from the extended source extraction has similar features to those of other comets, the shape is noticeably narrower than other comets \citep{harker2023dust}.

We converted the thermal emission spectrum to a relative emissivity spectrum by fitting the continuum in the 7.5--8.0 and 12.5--13.0 $\mu$m regions with a Planck function and dividing the observed spectrum by the fitted continuum \citep[e.g.,][]{kelley2017mid}. The fitted color temperature of $\sim$198~K is consistent with the 192~K color temperature reported by \cite{belyakov2026dust}. We consider the coma-corrected point source emissivity spectrum (Figure \ref{fig:spectra}, right) to be a more accurate reduction of 3I's MIRI LRS spectrum. Our custom reduction process significantly improved the shape of the critical 9--12 $\mu$m region.

\begin{figure*}[h]
\epsscale{0.7}
\plotone{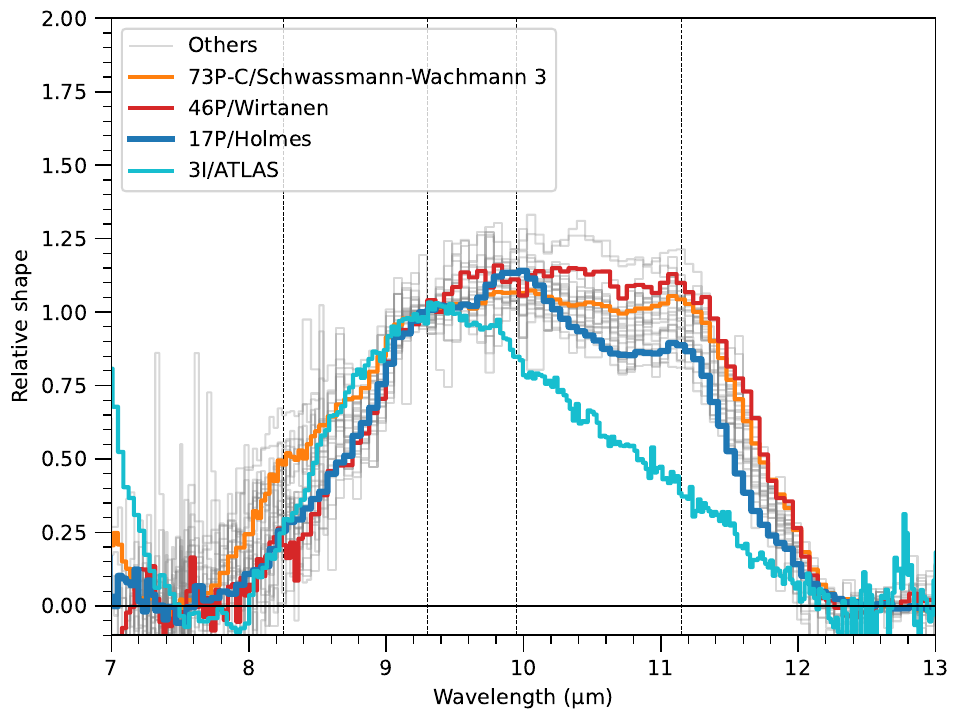}
\caption{Comparison of 10 $\mu$m silicate emission feature shapes, adapted from \cite{harker2023dust} to include 3I/ATLAS. The relative emissivity spectra are continuum subtracted and normalized to 1.0 at 9.3~$\mu$m. The distinct shape of the 3I silicate feature demonstrates its unusual mineralogy relative to Solar System comets. 17P/Holmes, 46P/Wirtanen, and 73P-C/Schwassmann–Wachmann 3 are highlighted and all other comet spectra with well-defined silicate bands are shown as thin gray lines.  A vertical line at 8.25~\micron{} marks the location of a feature in the 17P and 73P spectra, a line at 9.3~\micron{} marks the normalization point, lines at 9.95 and 11.15~\micron{} mark the locations of Mg-rich crystalline olivine features in the spectra of 17P and 73P.}
\label{fig:comparison}
\end{figure*}

The unusual shape of 3I's calibrated silicate emissivity feature is immediately apparent in comparison to the 10 $\mu$m silicate features observed in other Solar System comets (Figure \ref{fig:comparison}). The 3I feature is smooth and lacks the spectral structure associated with crystalline silicates. 

\section{Dust Composition Modeling} \label{models}

We focus our compositional analysis on the prominent 10 $\mu$m emissivity feature. This feature is commonly observed in cometary and asteroidal mid-infrared spectra and arises from Si-O vibrational modes in silicate minerals \citep[e.g.,][]{salisbury1991midinfrared, hanner19948}. Its shape, strength, and structure are diagnostic of the silicate mineralogy, grain size, morphology, and porosity \citep[e.g.,][]{emery2006thermal,min2005modeling,wooden2008cometary}.  
We explored a wide range of candidate dust compositional components drawn from published studies of Solar System comets in mid-infrared wavelengths \citep[e.g.,][]{kelley2017mid,harker2023dust,lisse2020spitzer}, including amorphous and crystalline silicates, sulfides, and amorphous carbon. Early models were evaluated by comparing their ability to reproduce the observed emissivity feature.

The 3I dust spectrum is consistent with a complete lack of crystalline silicates. We examined the spectral signatures of crystalline olivines and pyroxenes \citep{koike2003compositional,chihara2002compositional} and found them to be inconsistent with the observed spectrum. Even small amounts of crystalline material produce distinct spectral features that are not present in the observed spectrum. The measured feature is broad and smooth, lacking the sharp structure typically associated with crystalline olivine or pyroxene. Consequently, we exclude crystalline silicates from our detailed compositional modeling. We discuss formal constraints on the abundance of crystalline silicates in Section \ref{sec:results}.
Troilite was also included in the preliminary examination since iron sulfides have been identified in cometary dust, including in comet 1P/Halley and returned from comet 81P/Wild 2 \citep{jessberger1988aspects,zolensky2006mineralogy}. Models containing troilite introduced spectral structure that degraded the fit compared to silicate-carbon mixtures. Therefore, we conclude that sulfides, including troilite, are not required to model the emissivity feature. 
After this initial compositional exploration, our model was restricted to amorphous carbon, amorphous olivine, and amorphous pyroxene \citep{edoh1983optical,dorschner1995steps}.  The amorphous silicates are not true minerals, but they represent dust with an average stoichiometry similar to pyroxene, (Mg$_x$,Fe$_{1 -x}$)SiO$_3$, or olivine, (Mg$_x$,Fe$_{1-x}$)$_2$SiO$_4$.

To model the continuum-removed MIRI emissivity spectrum, we constructed a multi-dimensional grid of single grain thermal emission models using the {\tt grains2} dust thermal emission package \citep{michael_kelley_2026_18516737}. The {\tt grains2} model follows the approach of \citet{harker2007di}, where mono-compositional spherical dust grains are assumed to be in local thermal equilibrium with absorbed sunlight at the distance of the comet. Grain porosity is modeled by mixing optical constants of the material with vacuum using Bruggeman's effective medium approximation \citep{bruggeman1935}, and grain absorption efficiencies are calculated with Mie theory.

The thermal emission for each grain was calculated over a grid of grain sizes, wavelengths, and grain porosities, with wavelength array defined by the calibrated MIRI LRS spectrum ($\geq$~5.2 $\mu$m). We fixed the observing geometry to those of the science observation: $r=2.214$ au and $\Delta=1.798$ au. Grain radii were sampled logarithmically for $a=0.1-1000$ $\mu$m. Porosity, $P$, of the grain was parameterized by $P=1 - (a/a_0)^{D-3}$, where $D$ is the grain fractal dimension and $a_0=0.1$~\micron{} is the monomer grain size.  The fractal dimension is allowed to vary from $D=2.5-3.0$, where $D=3$ corresponds to solid grains and $D=2.5$ corresponds to a fluffy aggregate within the limits of what Mie theory can reliably model. We include three compositional components: amorphous carbon \citep{edoh1983optical}, amorphous olivine, and amorphous pyroxene \citep{dorschner1995steps}. We selected the optical constants for amorphous olivine and pyroxene with compositions of 50\% Mg and 50\% Fe.

For each dust component, we used the Hanner differential size frequency distribution \citep{hanner1983nature},  $dn/da=(1 - a_0/a)^{-M}(a_0/a)^{-N}$, with large grain slope parameter ($N$), peak grain radius ($a_p=a_0(M+N)/N$), and maximum grain radius ($a_{max}$). Allowing a different peak grain size for carbonaceous and silicate dust provided remarkable improvements in the fit of the silicate band shape and the continuum temperature. Therefore, different peak radii were allowed for the silicate and carbon components, but all components shared the same $N$ and $a_{max}$. Each model spectrum was constructed as a weighted sum of carbon, olivine, and pyroxene emission. Model scale factors are based on the number of grains at the peak of the differential grain size distribution ($a_p$). The relative abundances were determined by the silicate-to-carbon and pyroxene-to-olivine mass ratios using bulk grain densities of 3.3~g~cm$^{-3}$ for silicates and 1.5~g~cm$^{-3}$ for amorphous carbon.

\begin{figure*}[ht!]
\plotone{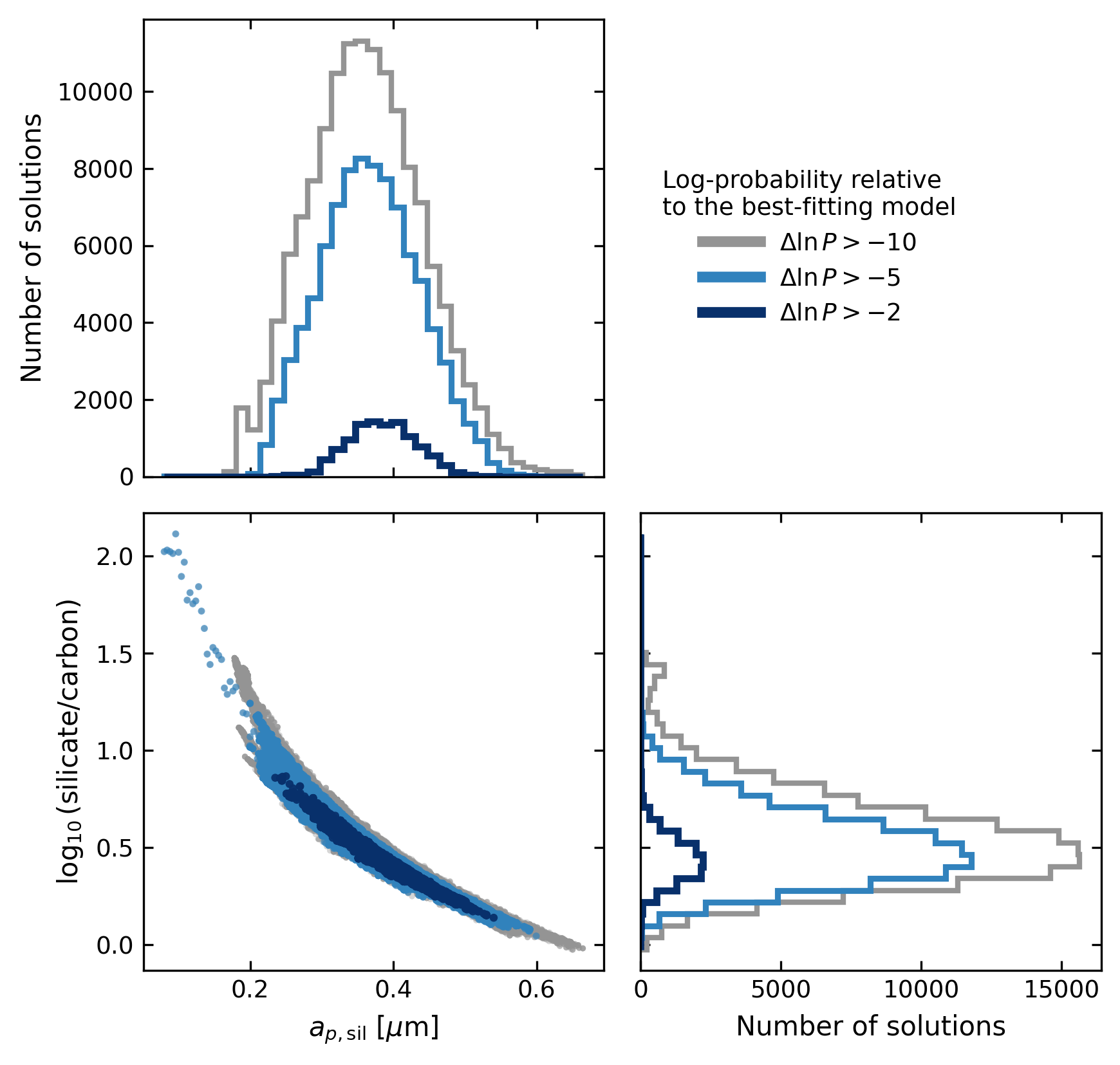}
\caption{Joint posterior distribution for $\log_{10}(\mathrm{sil}/\mathrm{car})$ versus $a_{p,\mathrm{sil}}$. $\Delta\ln P$ denotes the calculated difference from the maximum profile likelihood. The colors show solutions within $\Delta\ln P >$ -10, -5, and -2 of the best-fitting model. The relationship between silicate abundance and silicate peak grain size is clear across a space of high-probability solutions. }
\label{fig:corner}
\end{figure*}

To identify the dust properties consistent with our observed emissivity spectrum, we explored the parameter space defined by the model grid using the emcee Markov Chain Monte Carlo (MCMC) package \citep{foreman2013emcee}. We allowed interpolation between the grain radius and fractal dimension to enable the MCMC to smoothly sample the full parameter space. For each MCMC step, the single-grain thermal emission was taken from the grid and combined according to the Hanner size distribution and relative abundances of the compositional components to generate an emissivity spectrum. We fit the data over wavelengths from 5.6 to 13.2 $\mu$m. We adopted Gaussian uncertainties for each emissivity point of 0.025 for $\lambda <$ 12.7 $\mu$m, which are substantially larger than the formal extracted uncertainties from the JWST pipeline. This is a deliberately conservative uncertainty choice to account for systematic uncertainties in the reduction process and spectral shape that are not captured by the formal pipeline uncertainties. The change to the uncertainty reduces the weight on long-wavelength data outside the silicate emission feature, which is lower signal-to-noise. Each modeled emissivity spectrum was compared to the observed spectrum using a weighted $\chi^2$ likelihood.

Early MCMC exploration found that the outcome was sensitive to the distribution of the priors and that there was a clear degeneracy of solutions related to the silicate peak grain size and the relative amounts of silicates and carbon. To avoid biasing the resulting posteriors toward any particular solution, the final MCMC used a broad and uniform distribution of priors that included the full range of the model grid.  We initialized the MCMC with 96 walkers that was evolved for 25,000 steps to ensure that the chain had fully sampled the parameter space. We discarded the first 5,000 steps and did not thin the MCMC posteriors.

\section{Results} \label{sec:results}

Table \ref{tab:dust_model} presents the posterior distribution of our eight MCMC model parameters: $\log_{10}(\mathrm{sil}/\mathrm{car})$,
$a_{p,\mathrm{car}}$,
$a_{p,\mathrm{sil}}$,
$N$,
$\log_{10}(\mathrm{pyx}/\mathrm{ol})$,
$D_{\mathrm{car}}$,
$D_{\mathrm{sil}}$, and
$\log_{10}(a_{\max})$.
The posterior distribution exhibits a clear correlation between the silicate peak grain size and the relative abundance of silicates and carbon (Figure \ref{fig:corner}), indicating that these parameters are not independently constrained by the LRS spectrum. The MCMC results also favor a pyroxene-rich silicate composition, with $\log_{10}(\mathrm{pyx}/\mathrm{ol})=2.25^{+0.52}_{-0.60}$ or a pyroxene-to-olivine ratio of $\sim$180:1 for the grain size distribution scale factors. The fit does not effectively constrain the maximum grain size.  This is to be expected given the limited wavelength range of the LRS spectrum, since longer wavelength observations provide better sensitivity to the large-grain size distribution \citep[e.g.,][]{harker2011mid}. Additionally, the posterior distribution for the fractal dimension, $D$, reaches the upper boundary at $D=3$ for both carbon and silicate grains. This preference for the upper bound favors compact grains over highly porous aggregates. 

\begin{figure*}[ht!]
\plotone{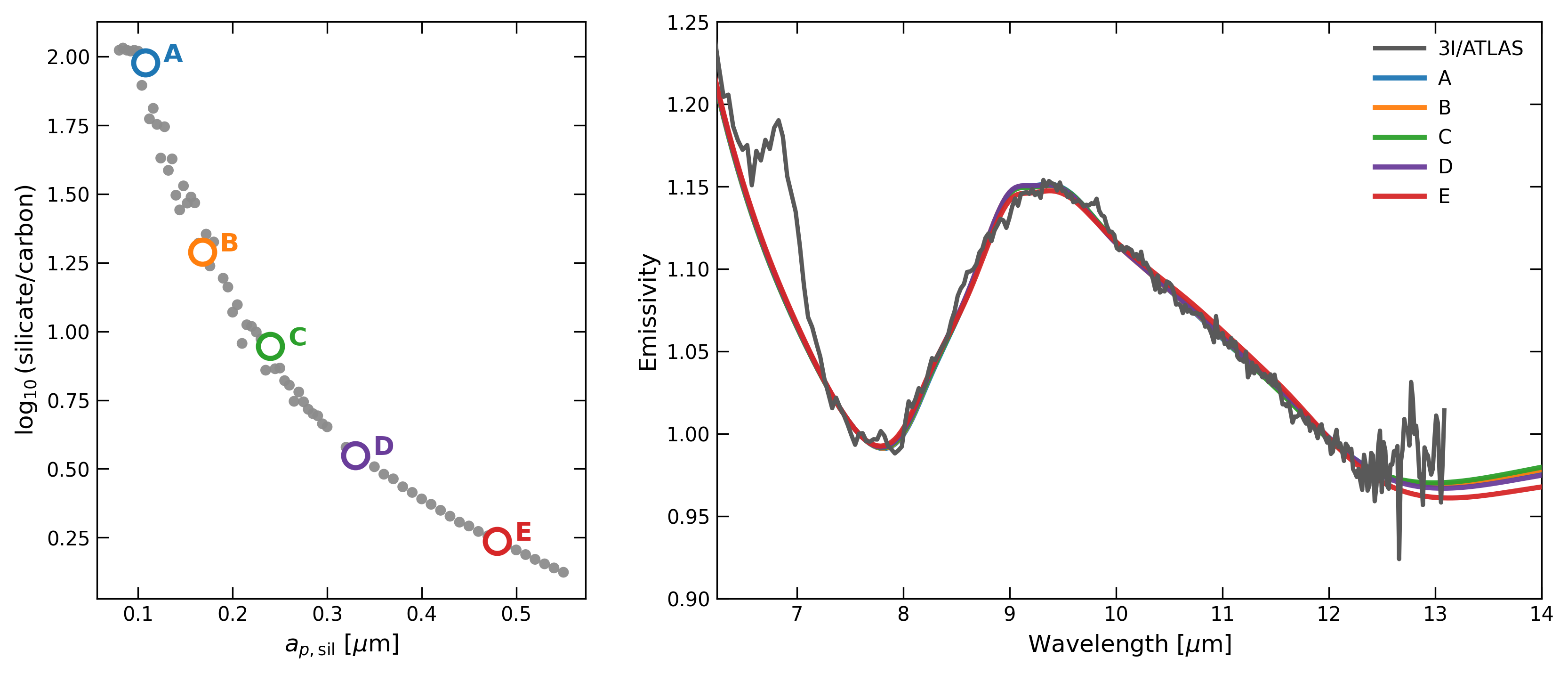}
\caption{Left: The calculated ridge of solutions through the $\log_{10}(\mathrm{sil}/\mathrm{car})$ versus $a_{p,\mathrm{sil}}$ parameter space. Solutions A-E correspond to the 10th, 30th, 50th, 70th,
and 90th percentiles in $a_{p,\rm sil}$ among solutions with
$\Delta\ln P>-5$. Right: 3I/ATLAS emissivity spectrum compared to the modeled spectra for solutions A-E. All solutions are reasonable fits to the silicate emissivity feature. Our dust model does not include the emission feature observed at 6.83 $\mu$m.}
\label{fig:models}
\end{figure*}

The degeneracy between $a_{p,\mathrm{sil}}$ and $\log_{10}(\mathrm{sil}/\mathrm{car})$ contains a continuum of dust models that all fit the emissivity spectrum well. We identify a space of plausible solutions for these two parameters by fixing $a_{p,\mathrm{sil}}$ at a series of values and optimizing the solution using a differential evolution algorithm. The resulting dataset defines a ridge of possible solutions through the parameter space. 
We demonstrate the effect of this particular degeneracy on the modeled spectra by selecting five representative solutions along the profile-likelihood ridge (Figure \ref{fig:models}, left). These solutions (A–E) correspond to the 10th, 30th, 50th, 70th, and 90th percentiles in $a_{p,\mathrm{sil}}$ space among the profile-likelihood solutions with $\Delta\ln P > -5$, where $\Delta\ln P$ is measured relative to the maximum profile likelihood. A limit of $\Delta\ln P > -5$ restricts the analysis to statistically plausible solutions that are near the optimal MCMC posteriors. The model parameters for each of the A--E solutions are given in Table \ref{tab:ridge_models}.  We also include calculated total mass and silicate-to-carbon mass ratios for the sub-micron grains and the total coma.  The silicate-to-carbon mass ratio depends on the size regime because we used separate grain size distributions for the carbonaceous and silicate dust.  For each solution, we compute a model emissivity spectrum using the {\tt grains2} package. Each of these spectra match the 10 $\mu$m emissivity feature and surrounding continuum (Figure \ref{fig:models}, right).

\begin{deluxetable}{lc}
\tablecaption{MCMC posterior medians for our eight parameters. We give uncertainties at the $1\sigma$ level. \label{tab:dust_model}}
\tablehead{
\colhead{Parameter} &
\colhead{Posterior Constraint}
}
\startdata
$\log_{10}(\mathrm{sil}/\mathrm{car})$
    & $0.52_{-0.19}^{+0.25}$ \\
$a_{p,\rm car}\,[\mu\mathrm{m}]$
    & $0.97_{-0.04}^{+0.03}$ \\
$a_{p,\rm sil}\,[\mu\mathrm{m}]$
    & $0.36_{-0.08}^{+0.08}$ \\
$N$
    & $-3.48_{-0.17}^{+0.22}$ \\
$\log_{10}(\mathrm{pyx}/\mathrm{ol})$
    & $2.25_{-0.60}^{+0.52}$ \\
$D_{\rm car}$
    & $2.99_{-0.02}^{+0.01}$ \\
$D_{\rm sil}$
    & $2.98_{-0.03}^{+0.01}$ \\
$\log_{10}(a_{\max})$
    & $1.16_{-0.20}^{+0.32}$ \\
\enddata
\tablecomments{Abundance ratios are by number of grains at the peak of the size distributions ($a_p$).}

\end{deluxetable}

\begin{deluxetable}{lcccccc}
\tablecaption{Parameter values for representative models A-E. These models are selected from the profile-likelihood ridge at the 10th, 30th, 50th, 70th,
and 90th percentiles in $a_{p,\rm sil}$ among solutions with
$\Delta\ln P>-5$. \label{tab:ridge_models}}
\tablehead{
\colhead{Parameter} &
\colhead{Unit} &
\colhead{A} &
\colhead{B} &
\colhead{C} &
\colhead{D} &
\colhead{E}
}
\startdata
$\log_{10}(\mathrm{sil}/\mathrm{car})$ & & 1.978 & 1.289 & 0.946 & 0.549 & 0.238 \\
$a_{p,\rm car}$              & \micron{} & 0.976 & 0.971 & 0.950 & 0.980 & 0.986 \\
$a_{p,\rm sil}$              & \micron{} & 0.108 & 0.168 & 0.240 & 0.330 & 0.480 \\
$N$                                    & & -2.589 & -2.861 & -3.242 & -3.279 & -3.662 \\
$\log_{10}(\mathrm{pyx}/\mathrm{ol})$  & & 1.573 & 2.242 & 2.362 & 2.097 & 2.967 \\
$D_{\rm car}$                          & & 3.000 & 3.000 & 3.000 & 3.000 & 3.000 \\
$D_{\rm sil}$                          & & 2.950 & 2.950 & 2.950 & 3.000 & 3.000 \\
$\log_{10}(a_{\max})$        & \micron{} & 0.777 & 0.846 & 1.043 & 0.958 & 1.161 \\
$\Delta\ln P$                          & & -4.05 & -3.04 & -2.21 & -0.51 & -0.58 \\\hline
$m_1$\tablenotemark{a}          & $10^3$ kg & 11.4 & 11.9 & 12.3 & 11.9 & 11.1 \\
$m_1$(sil/car)\tablenotemark{a}           & & 3.07 & 3.20 & 3.31 & 3.14 & 2.82 \\
$m_\mathrm{T}$\tablenotemark{b} & $10^3$ kg &  688 &  772 & 1038 &  926 & 1178 \\
$m_\mathrm{T}$(sil/car)\tablenotemark{b}  & & 0.19 & 0.19 & 0.18 & 0.27 & 0.33 \\
\enddata
\tablenotetext{a}{Derived sub-micron mass and sub-micron silicate-to-carbon mass ratio.}
\tablenotetext{b}{Derived total mass and total silicate-to-carbon mass ratio.}
\end{deluxetable}

The absence of clear crystalline silicate features in the MIRI spectrum suggests that only a small fraction of the silicates could be crystalline. We use the five representative models (A–E) to calculate an upper limit on the crystalline fraction within the observed spectrum. We introduce crystalline forsterite and enstatite by replacing a corresponding fraction of the amorphous silicate emission meaning that the resulting crystalline upper limit represents a fraction of the silicate emission, not a silicate mass fraction.

We define a baseline $\chi^2$ between the amorphous model spectrum and the observed emissivity spectrum over the observed spectrum (5.6–13.2 $\mu$m) with a mask at 6.6-7.1 $\mu$m so that the 6.8 $\mu$m feature does not impact the calculation. We investigate the crystalline upper limit with two approaches. First, we use a conditional analysis where we hold all eight dust parameters fixed and replace amorphous silicates with crystalline silicates. In this case, we allow the crystalline material to be distributed between olivine and pyroxene and identify the combination that produced the minimum $\chi^2$. Then, we use a fully profiled analysis where we start with the same models and allow all eight dust parameters and the mixture of crystalline material to vary at each crystalline fraction.

For both analyses, we define the upper limit on the crystalline fraction for each model as the value at which $\Delta\chi^2$ reaches the 99\% $\chi^2$ threshold, corresponding to a threshold of $\Delta\chi^2 = 6.63$ \citep{press1992numerical} where $\Delta\chi^2=\chi^2(f_{\rm cryst})-\chi^2(f_{\rm cryst=0})$. We determine the crystalline upper limit by linearly interpolating between the adjacent grid points that bracket the $\Delta\chi^2=6.63$ threshold. The baseline $\chi^2(f_{\rm cryst=0})$ for models A-E range from 89.9 to 93.3 with the reduced $\chi^2$ ranging from 0.32-0.33. For the fully profiled analysis, we optimized model E, which had the lowest $\chi^2$ of the representative models. 
We use the conservative emissivity uncertainty adopted for the MCMC analysis. 

The conditional analysis results in 99\% upper limits of 2.3-4.1\% across models A-E. For the fully profiled analysis, all dust parameters in each model were allowed to vary resulting in an increase of upper limit to 5.3\%. The minimum $\chi^2$ for all models occurs at zero crystalline abundance, which means that the spectrum does not require crystalline material. 


In addition to the primary silicate feature, there is another emission feature that peaks at approximately 6.83 $\mu$m that is not modeled (as seen in Figure \ref{fig:models}). We measured the peak location by subtracting the dust model emissivity spectrum from our observed spectrum. Our measurement of the feature's peak wavelength is limited by the spectral resolution of the LRS mode, which has a median sampling of 0.04 $\mu$m. We do not include this feature in our silicate dust modeling, but it does shed light on other compositional components present in 3I's coma. Higher-resolution MIRI MRS observations provide a more detailed characterization of this feature \citep{belyakov2026dust}. 

\cite{belyakov2026dust} favor carbonate minerals as the origin of this feature based on their MIRI MRS spectrum, although they acknowledge aliphatic hydrocarbons as a possible explanation. Carbonates have been identified or inferred in cometary dust, including in material returned from 81P/Wild 2 \citep{mikouchi2007mineralogy}. 
Aliphatic C-H bearing material is another possible explanation for the feature, exhibiting near- and mid-infrared bands at 3.4, 6.85, and 7.25 $\mu$m in the laboratory \citep{gadallah2012mid} and the interstellar medium \citep{chiar2000composition}. Similar aliphatic C–H features are also observed in laboratory spectra of meteoritic insoluble organic matter \citep{orthous2013mid,kaplan2019reflectance}. This interpretation is consistent with the C–H stretching residuals reported at 3.4~$\mu$m in JUICE MAJIS observations of 3I at $r_h=1.36-1.68$~au \citep{bockelee2026near}. No such residuals were identified in later, higher-resolution NIRSpec observations at $r_h=2.4$~au \citep{roth2026isotopic}, although uncertainties in the CH$_3$OH emission model complicate the interpretation of the 3.4~$\mu$m region. The different heliocentric distances may also contribute to the apparent discrepancy due to temporal evolution of the coma dust, including a possible temperature dependence of the 3.4 $\mu$m C-H emission feature \citep{chyba1989heliocentric}. We do not detect the 7.25 $\mu$m band in our spectrum. However, the strength of the 7.25 $\mu$m band is expected to be substantially weaker than that of the observed 6.83~\micron{} band and, consequently, is less frequently detected in astronomical spectra \citep{yang2016aliphatic}.
A similar feature at 6.85 $\mu$m was identified in the MIRI MRS spectrum of comet C/2017 K2 by \cite{woodward2025jwst}, who attributed it to either aliphatic hydrocarbons or hydrogenated Polycyclic Aromatic Hydrocarbons (PAHs). 
The 6.83~\micron{} feature is consistent with both carbonate minerals and aliphatic C–H bearing material. Aliphatic hydrocarbons provide a plausible explanation based on laboratory spectra and observations of other comets, but our LRS observations do not distinguish between these possible interpretations.

\section{Discussion} \label{sec:discussion}

The lack of crystalline silicates demonstrated by our analysis suggests that the dust component of the 3I/ATLAS nucleus underwent little thermal processing during its formation and accretion. These findings are consistent with the isotopic and volatile compositions determined from complementary JWST and ground-based spectroscopy, demonstrating that the object has preserved an unusually high HDO/H$_2$O ratio \citep{cordiner2026isotopic}, large hypervolatile abundances \citep[i.e., CO;][]{biver2026perihelion}, and a high CH$_4$ and CH$_3$D content \citep{roth2026isotopic}, indicative of extremely cold formation conditions. \cite{furuya2026high} find that the exceptionally high observed D/H ratio in water is most readily produced in a low metallicity interstellar/protostellar environment, and can only be preserved if the material experiences limited subsequent thermal processing. The preservation of these isotopic and molecular signatures implies that 3I/ATLAS experienced minimal heating and limited radial transport and processing within its stellar disk, both of which allowed the amorphous silicates and large abundances of extremely volatile ices to remain intact. That 3I/ATLAS formed in a relatively quiescent part of its natal disk, subject to reduced radial mixing, provides another point of contrast with our own Solar System comets, which incorporated a mixture of materials, including those processed at high temperatures, at relatively small distances from the proto-Sun \citep{brownlee2006}.

\citet{belyakov2026dust} obtained a 5--28~\micron{} spectrum of 3I/ATLAS with JWST's MIRI instrument in the Medium Resolution Spectroscopy mode ($R\sim3000$).  They fit the 7.5--13.5~\micron{} feature using a model based on the agglomerated debris model of \citet{zubko2009}, and find an upper-limit of 3.1\% to the crystalline silicate mass fraction, similar to our limit of 5.3\%.  They also modeled the full spectrum with the ``DI'' dust model of \citet{lisse2006}, finding amorphous olivine-type, amorphous carbon dust, and carbonate dust (via dolomite), but also other species in minor amounts: Mg, Fe, and Ca-Mg crystalline silicates, and smectite, a phyllosilicate.  This is in contrast to our results, which nearly exclude amorphous olivine-type dust in favor of pyroxene-type dust.  We did not formally test for minor species, but note that they were not required for a good fit to the LRS spectrum.

Our models of the silicates in 3I/ATLAS also show a strong preference for the presence of pyroxene-type over olivine-type dust. The relative abundances of these silicates can be sensitive to the refractory elemental composition of the material from which the object formed. For example, equilibrium condensation calculations from \citet{jorge2022forming} show that decreasing Mg/Si can shift mineralogy from forsterite (Mg-rich olivine endmember) to enstatite (Mg-rich pyroxene endmember). 
This relationship is particularly interesting given the observational evidence that 3I formed in a relatively metal-poor environment. 
We note that our amorphous olivine and pyroxene optical constants contain equal proportions of Mg and Fe. 

Interestingly, the dust components favored here—submicron amorphous pyroxene-type silicates and larger amorphous carbon grains—are broadly consistent with those inferred from near-infrared JWST/NIRSpec observations of the coma using the same optical constants \citep{protopapa2026jwst}. In that analysis, the pre-perihelion spectrum is best reproduced by two populations of aggregates: smaller water-ice-bearing grains containing amorphous pyroxene (favored over olivine), and a coarser population containing amorphous pyroxene and amorphous carbon. Although the two analyses probe different wavelength regimes and are sensitive to potentially different grain populations, they suggest similar refractory constituents in the coma of 3I/ATLAS. Furthermore, the pre-perihelion NIRSpec spectrum shows weak residual structure near the aliphatic C–H stretching bands at approximately 3.4–3.5 µm, longward of organic gas emission \citep{protopapa2026jwst}. Although the identification is tentative, this structure may be consistent with the aliphatic C–H-bearing material proposed here to explain the 6.83 µm feature.

Previous work has presented many thermal emission models of cometary spectra, but the details of the models are different.  Our model is closest to that used by \citet{harker2023dust}, but with one important distinction.  \citet{harker2023dust} assumed all amorphous dust species (pyroxene-type, olivine-type, and amorphous carbon) had the same grain size distribution.  Here we required different grain size distributions for the silicate and carbonaceous grains, and ultimately found that the silicate dust was dominated by amorphous pyroxene-type grains.  The results of \citet{harker2023dust}, based on model fits to 57 spectra of 33 comets, also tend to favor amorphous pyroxene-type over olivine-type dust when a similar wavelength range (5--13~\micron) is considered.  Taking the caveat that the free parameters of the models are different, it appears the difference between 3I and solar system comets primarily rests in the presence of Mg-rich crystalline silicate dust in the latter.  Similarly, the surfaces of centaurs and trans-Neptunian objects also show evidence for significant amounts of crystalline silicates \citep{vernazza2025-silicates}.

Regarding the amorphous silicate content, models of the dust in the diffuse interstellar medium suggest it is dominated by amorphous olivine-type material \citep{kemper2004} or even intermediate (non-stoichiometric) material \citep{fogerty2016-ism}, rather than the pyroxene-type material in 3I/ATLAS and solar system comets.  A similar finding was presented by \citet{demyk2000}, who found that the dust around evolved stars tended to be olivine-type, yet the dust around protostars was pyroxene-type and/or aluminosilicate.  They suggested that an unknown process may convert the dust before it is incorporated into molecular clouds.  3I's amorphous dominated feature presents the clearest evidence that this evolved dust may be preserved within cometary objects, but how this evolution occurs early in the galaxy's evolution should be more carefully considered.

While some studies suggest that galactic cosmic rays can convert crystalline silicates to amorphous \citep{bringa2007energetic}, this process requires energetic heavy-ion irradiation. Models of cosmic ray irradiation predict significant modification of cometary ice at depths up to $\sim$15–20 m \citep{maggiolo2026interstellar}, but the depth at which silicates can be amorphized has not been similarly constrained. It may be shallower than the depth of ice modification due to the fact that ice modification occurs through a broader range of processes. Models of 3I's volatile production rates suggest that it lost $\gtrsim$10 m of surface material during its passage through our Solar System \citep{frincke2026assessment}. This amount of erosion may have exposed less processed interior layers, suggesting that the dust observed by JWST after perihelion was more representative of its primordial composition. 

\section{Conclusions} \label{sec:conclusion}

We present JWST/MIRI LRS spectroscopy of 3I/ATLAS to investigate the dust composition of the target during our post-perihelion observation. The smooth shape of our observed 10 $\mu$m silicate emissivity feature is distinctly different from those commonly observed in Solar System comets. Compositional modeling shows the object is dominated by amorphous pyroxene silicate composition and the wavelengths available in our observations result in a degeneracy between the peak silicate grain size and the silicate/carbon abundance. Our set of representative models all successfully reproduce the silicate feature. Our spectrum requires no crystalline silicates and using a conservative uncertainty estimate we place an upper limit on the amount of crystalline silicate material at 5.3\%. We attribute a separate emission feature at $\sim$6.83 $\mu$m to aliphatic C-H material. The amorphous silicate composition, along with previous publications regarding 3I's high D/H and hypervolatile inventory suggest material that has experienced little thermal processing during its history. 

\begin{acknowledgments}
This work is based on observations made with the NASA/ESA/CSA James Webb Space Telescope. The data were obtained from the Mikulski Archive for Space Telescopes at the Space Telescope Science Institute, which is operated by the Association of Universities for Research in Astronomy, Inc., under NASA contract NAS 5-03127 for JWST. 

Support for program GO-5094 was provided by NASA through a grant from the Space Telescope Science Institute, which is operated by the Association of Universities for Research in Astronomy, Inc., under NASA contract NAS 5-03127.

The authors thank Sarah Kendrew and Ian Wong (STScI) for assistance with the MIRI LRS calibration issues that we encountered. C.A.T. also thanks Joshua Emery for helpful discussions about silicates in mid-infrared wavelengths. 

Part of this research was conducted at the Jet Propulsion Laboratory, California Institute of Technology, under a contract with the National Aeronautics and Space Administration (80NM0018D0004).

MND’s work is funded by the European Union. Views and opinions expressed are however those of the author(s) only and do not necessarily reflect those of the European Union or the European Research Council Executive Agency. Neither the European Union nor
the granting authority can be held responsible for them. MND’s work is supported by ERC grant PSII
(DOI: 10.3030/101230593).

\end{acknowledgments}

\begin{contribution}

C.A.T. led the MIRI LRS data reduction, analysis, dust modeling, interpretation, and preparation of the manuscript. M.S.P.K. developed the grains2 dust thermal emission software, defined the initial model grid, developed the extended-source slit-throughput correction, and calculated dust masses. D.E.H. contributed expertise in cometary dust thermal modeling and verified certain aspects of the modeling methodology and results. M.A.C. was Principal Investigator of the JWST program. M.S.P.K., D.E.H., and M.A.C. contributed to the interpretation of the modeling results and preparation of the manuscript. D.F. computed the ephemeris of 3I to ensure a successful detection by JWST. 

D.B., S.B.C, M.N.D., M.M., S.N.M., C.O., S.P., N.X.R., and M.E.S. contributed to scientific discussion and interpretation of the results for this work and the overall work for this 3I program. They also reviewed and edited the manuscript.


\end{contribution}

%
\facilities{JWST(MIRI)}

\software{\texttt{astropy} \citep{astropy2013, astropy2018, astropy2022}, \texttt{jwst} \citep{bushouse2024}, \texttt{matplotlib} \citep{hunter2007}, \texttt{numpy} \citep{harris2020}, \texttt{grains2} \citep{michael_kelley_2026_18516737}, \texttt{mskpy} \citep{Kelley2021mskpy}, \texttt{emcee} \citep{foreman2013emcee}, and \texttt{scipy} \citep{virtanen2020}.}



\bibliography{3I_miri_lrs}{}
\bibliographystyle{aasjournalv7}



\end{document}